# A BRIEF HISTORY OF DARK ENERGY

C Sivaram

Indian Institute of Astrophysics, Bangalore- 560034

Gurzadyan-Xue Dark Energy was derived in 1986 (twenty years before the paper of Gurzadyan-Xue). The paper by the present author, titled "The Planck Length as a Cosmological Constant", published in Astrophysics Space Science, Vol. 127, p.133-137, **1986**[1] contains the formula claimed to have been derived by Gurzadyan-Xue (in 2003)! Many recent papers, for example H J Mosquera Cuesta, et al, 2008[2], give the formula as:

$$\rho_{GX} = \frac{\pi}{8}\frac{c^4}{G}\frac{1}{a^2} \qquad \ldots (1)$$

This is identical to the one in the earlier mentioned paper of 1986, which has a detailed derivation based on the higher curvature power expansion for the vacuum fluctuation energy.

The above mentioned paper (C Sivaram, 1986) has in equations (2) – (6), just the above dark energy formula. Equation (2) has:

$$\rho_{vac} = \frac{1}{2}\hbar c R N^2 k_0^2 \qquad \ldots (2)$$

Where, $R \approx R_H^{-2}$ (or '$a$' the scale factor), $N^2 k_0^2$ shown to be the Plank wave number $\frac{c^3}{\hbar G}$, so equation (2) just gives:

$$\rho_{vac} = \frac{1}{2}\frac{c^4}{G}\frac{1}{R_H^2} \qquad \ldots (3)$$

Again the dark energy $\rho_{vac}$ has been evaluated as $10^{-8} ergs/cm^3$ or $10^{-29} gm/cm^3$, exactly what is now implied by the present observations.

Indeed the above paper (C Sivaram, 1986) has other related papers referred to, by the same author, published in 1986 and earlier.

For example, Uncertainty Principle Limits on the Cosmological Constant, Int. J. Theor. Phys., 1986[3] also gives the same value for the vacuum energy.

Earlier papers (C Sivaram, 1986; 1982[4,5]) also expresses the dark energy in terms of fundamental constants.

Indeed the latter, published in 1982, gives the Hubble constant and consequently age of a vacuum dominated universe as:

$$t_U = \frac{g^4}{Ge^8}\left(\frac{c^5 G_F^3}{\hbar}\right)^{1/2} = 14 \, \text{Gyr} \qquad \ldots (4)$$

(Which agrees with the W-MAP, i.e., 13.7 Gyr!)

The constants $g, e, G_F, G$ are the strong, electromagnetic, weak and gravitational couplings!

Another paper (C Sivaram, 1994[6]) predicts a 0.8 domination of the universe density by vacuum energy (based on the large scale dynamics).

In their recent note arxiv:astro-ph/0510459v1[7], Gurzadyan and Xue have stated that the formula for dark energy derived by Padmanabhan in Cl. Q. Grav., 2005 and expressed as $\rho_{vac} = L_{Pl}^{-2} L_H^{-2}$ was derived by them four years earlier. The above formula is just that derived in C Sivaram, 1986.

Even their equations (3) and (4), that is:

$\rho_\Lambda \approx \frac{hc}{a^n} N_{max}(N_{max}+1)$ and $N_{max} \approx \frac{a}{L_P} = 10^{61}$ is just that which was obtained in equations (5) and (6) of the above mentioned paper of 1986 on page <u>135</u>!

It is only fair, especially when the title Gurzadyan-Xue dark energy is used in several papers, that the above references of the present author be at least given due mention as it was derived more than 20 years earlier, when dark energy was not even remotely thought to dominate the universe.